\documentclass[aps,twocolumn,superscriptaddress]{revtex4}
\usepackage{mathrsfs,amssymb,amsbsy,graphicx}

\def\scr{\mathscr}
\def\CL{{\scr L}} \def\CH{{\scr H}}  
\def\CC{{\scr C}} \def\CD{{\scr D}}  
   
 \def\CS{{\scr S}} 

\def\Bid{{\mathchoice {\rm {1\mskip-4.5mu l}} {\rm
{1\mskip-4.5mu l}} {\rm {1\mskip-3.8mu l}} {\rm {1\mskip-4.3mu l}}}}

\def\avg#1{\langle#1\rangle}	\def\<{\langle} 	\def\>{\rangle} 
 		
\def\Tr{\textrm{Tr}}	\def\tr{{\rm tr}}
\def\pt{\partial}
\def\al{\alpha}		\def\bt{\beta}		
\def\sig{\sigma}	\def\del{\delta}	\def\Del{\Delta}
\def\eps{\epsilon}		\def\vphi{\varphi}
\def\up{\uparrow}	\def\down{\downarrow}

\def\Vk{{\mathbf k}}   
	\def\VD{\boldsymbol{D}}
  \def\V0{{\mathbf 0}}
\def\Vr{{\bf r}}
\def\VS{{\bf S}}

\renewcommand{\subparagraph}[1]{
{\bf #1}}

\def\be{\begin{equation}}	\def\ee{\end{equation}}
\def\bea{\begin{eqnarray}}	\def\eea{\end{eqnarray}}
\def\nn{\nonumber}

\def\suspin{SU($2$)$_{\rm spin}$}
\def\sucolor{SU($2$)$_{\rm color}$}
\def\uch{U(1)$_{\rm charge}$}

\begin{document}

\title{Spin-disordered superfluid state for spin-1 bosons with fractional
spin and statistics }

\author{W. Vincent Liu}
\affiliation{Center for Theoretical Physics, 
Department of Physics, Massachusetts Institute of
Technology, Cambridge, Massachusetts 02139}
\affiliation{Department of Physics, Massachusetts Institute of
Technology, Cambridge, Massachusetts 02139}

\author{Xiao-Gang Wen}
\affiliation{Department of Physics, Massachusetts Institute of
Technology, Cambridge, Massachusetts 02139}

\date{\today}
\begin{abstract}
We study a strongly correlated spin-1 Bose gas in 2D space by using the
projective construction. A spin-disordered superfluid state is constructed and
proposed as a candidate competing with the conventional polar condensate when
interaction is antiferromagnetic. This novel state has a non-trivial
topological order whose low energy excitations carry fractional spin, charge,
and statistics. The spin excitations become gapless only at the edge and are
described by level-1 \suspin\ Kac-Moody algebra. The edge state is identical
to the edge state of the chiral spin liquid or the right moving sector of
spin-1/2 chain.
\end{abstract}
\maketitle

So far most studies of  ultra cold alkali atomic gases
have been focused on Bose-Einstein condensation (BEC) in 
weakly interacting dilute limit
(for review, see for instance
\cite{Dalfovo:99}). By contrast, a strongly correlated Bose gas may
lead us to interesting novel phenomena.  While it is unclear whether
one can achieve a {\it stable} strongly correlated gas in 3D by simply
cranking up the scattering length without collapsing the system, a 2D
gas with even relatively weak interaction can be strongly correlated in
nature.  The argument is based on 
the renormalization group analysis \cite{Fisher:88} that showed 
that the interaction in 2D is
marginally irrelevant only in a dilute limit specified by $\ln \ln
(1/na^2) \gg 1$, where $n$ is the particle density and $a$ can be
thought of the interaction range or the scattering length.  The double
logarithm imposes a much stronger condition of the validity of the
dilute limit in 2D than the familiar $na^3\ll 1$ in 3D.    
For generic BEC system such as the disk condensate of
$^{23}$Na of Ref.~\cite{Ketterle:01pre} or a microelectronic chip of
condensed $^{87}$Rb atoms \cite{Hansel:01}, $\ln \ln (1/na^2)$ is of
$O(1)$.  These systems are, at least, not weakly correlated. Now the
question is that what is the ground state for those not-weakly correlated 2D
boson gas? In this letter, we shall 
show that a new class of 2D superfluid
can emerge due to the strong correlation.

\paragraph{Proposed spin-disordered superfluid.}
Let us consider a strongly correlated gas of spin-1 bosons
$\phi_m$ ($m=0,\pm1$) in a 2D homogeneous space with generic two-body
interactions  
$
H^{\rm int} =
\int d^2\Vr d^2\Vr^\prime \sum_{F=0}^2\CH^{\rm int}_F$
with \cite{Ohmi:98,Ho-Yip:98+00,Law:98}
\be \textstyle
\CH^{\rm int}_F = {1\over 2} V_F(\Vr-\Vr^\prime) C^{SS_z}_{mm^\prime}
C^{SS_z}_{n^\prime n} \phi^\dag_{m}(\Vr)
\phi^\dag_{m^\prime}(\Vr^\prime)
\phi_{n^\prime}(\Vr^\prime)\phi_{n}(\Vr)
\label{eq:Hboson}
\ee
where $C^{SS_z}_{m_1m_2} = \avg{SS_z | 11;m_1m_2}$ are the Clebsch-Gordan
coefficient for total spin $\VS=\bf{1}+\bf{1}$. In the alkali atomic
gases, the interactions for all three spin channels $V_S$, $S=0,1,2$ are
commonly short-ranged.
The Hamiltonian has a phase and spin rotation symmetry: \uch
$\times$ \suspin.  The  Hamiltonian 
derived by 
Ho \cite{Ho-Yip:98+00},
$\CH^{\rm int}={1\over 2}(c_0\hat{n}^2+c_2\hat{\VS}^2)$ where
$\hat{n}$ and $\hat{\VS}$ are the density and spin operators, 
corresponds to a special form of (\ref{eq:Hboson})
with $V_0(\Vr-\Vr^\prime)=(c_0-2c_2)\del(\Vr-\Vr^\prime)$,
$V_1(\Vr-\Vr^\prime)=(c_0-c_2)\del(\Vr-\Vr^\prime)$, and
$V_2(\Vr-\Vr^\prime) =(c_0+c_2)\del(\Vr-\Vr^\prime)$.
The case of antiferromagnetic spin interaction 
($V_2>V_0$ or correspondingly $c_2>0$) is somewhat intriguing. 
Mean field theories \cite{Ho-Yip:98+00,Law:98}
realize that the (coherent or fragmented) polar condensate
is energetically favored for this case. 
We note that the polar condensate, no matter coherent or 
fragmented, is a spin ordered state, as it directly condenses single
bosons ($\avg{\phi_{m}}\neq 0$) and therefore always 
breaks \suspin symmetry in the thermodynamical limit. 
For spin-1 bosons in an
optical lattice, Demler and Zhou speculated possible spin disordered
(liquid) 
phases and fractionalized spin excitations for arbitrary dimensions
\cite{Demler-Zhou:01pre}. 
But, a question remains: Is there a spin-disordered superfluid
competing with the  polar condensate? 

\begin{figure}[htbp]
\begin{center} 
\includegraphics[width=0.8\linewidth]{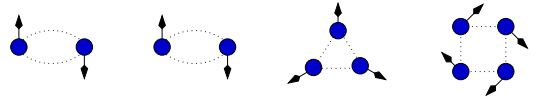}\vspace{-20pt}
\end{center}
\caption{A cluster configuration for $N=11$ spin-1 bosons with arrows
indicating spin. `Dotted line'$ = g(\Vr_i-\Vr_j)$.}
\label{fig:cluster}
\end{figure}

We have found such a spin-disordered state. 
It can be visulized as a gas of closed
(connected) clusters constituted by  various number of 
bosons; the size $l$ can take up the value
of $2$ (dimer), $3$ (triangle), $\ldots$, up to $N$ for an $N$ boson
system (see Fig.~\ref{fig:cluster}). 
Each cluster is a
spin singlet and is totally symmetric among all bosons within the
cluster. Mathematically, an $l$-cluster is described by
$C_l(\{\Vr_i,m_i\}_{i=1\ldots l}) = \prod_{i=1}^l g(\Vr_i-\Vr_{i+1})
\, \tr\{ \prod_{i=1}^l \eps \,  \CC^{m_i}\} $ (with $\Vr_{l+1} \equiv \Vr_1$
identified), 
where $g(\Vr)$ is a pairing wavefunction of $p_{x+iy}$-symmetry (thus
{\em odd} in $\Vr$), $\eps$ is a $2\times2$ antisymmetric matrix, and
$\CC^{m}$ is a (symmetric) $2\times2$ matrix defined by the
Clebsch-Gordan coefficients for ${1\over2}\otimes {1\over2}$, 
$\CC^m_{\al\bt} \equiv
\avg{(SS_z)1m|{1\over 2}{1\over 2}; \al\bt}$. 
The state 
is a superposition of all possible cluster configurations specified by
$\{n_l\}$ with $n_l$ the number of $l$-clusters. 
The wavefunction for the component of the state with $N$ bosons 
is, 
apart from an overall factor,
\bea
\Psi^b_N(\{\Vr_i,m_i\}_{i=1}^N) &\sim& \textstyle
\CS \sum_{\{n_l\}} 2^{N_c}
\prod_{p=1}^{N_c}  (l_p-1)!
\nn \\
&& \textstyle
 \quad \times  C_{l_p}(\{\Vr^p_{I}, m^p_{I}\}_{I=1}^{l_p}) \,,  
\label{eq:Psib=cluster}
\eea
where $N_c=\sum_l n_l$ is the total number of clusters, 
$p$ is a label used to sort 
all $N_c$ clusters, and $l_p$ is the size of the $p$-th cluster. 
In this labelling, the boson spins and coordinates
$\{\Vr_i,m_i\}_{i=1}^N$ are ono-to-one mapped onto the set
$\{\Vr^p_{I},m^p_{I}; p=1..N_c; I=1..l_p \}$.
The wavefunction is symmetrized for $N$ bosons, as indicated by 
$\CS$.  

Table.~\ref{tab:energy} compares the interaction energies
of the two states. 
\begin{table}[ht]
\begin{tabular}{l|c|c}
\hline
& \kern 1cm Polar \kern 1cm
 &\kern 1cm Spin-disordered \kern 1cm
\\ \hline
$\avg{\CH^{\rm int}_{S=0}}$ & ${1\over 6}V_0(\del\Vr) n_0^2$
&${1\over 6}V_0(\del \Vr) n_0^2+ O(|\del\Vr|^4)  $ \\ \hline
$\avg{\CH^{\rm int}_{S=1}}$ &  $0$  &  
$- (\#) {\pi\over 3}V_1(\del \Vr)n_0^4 |\del \Vr|^4$ \\ \hline
$\avg{\CH^{\rm int}_{S=2}}$ & ${1\over 3}V_2(\del\Vr) n_0^2$ 
& ${5\pi^2\over 9} V_2(\del \Vr) n_0^4 |\del \Vr|^4$ \\ \hline
\end{tabular}
\caption{Interaction energies calculated at mean field
level. Notations: $n_0$ is the density of condensed bosons;
$\del\Vr\equiv \Vr-\Vr^\prime$; $V_S(\del\Vr)$'s are
defined in Eq.~(\ref{eq:Hboson}). Note that 
the last two interaction energies for the spin-disordered 
vanish for $\del \Vr\rightarrow 0$.
In our calculation, the fermion representation (\ref{eq:phi=}) is
normalized by requiring $\avg{\phi^\dag_m\phi_m}$ equal to the 
boson density.
}
\label{tab:energy}
\end{table}
One sees that the spin-disordered state we proposed has a lower
interaction energy when $V_2$ is positively large and short ranged.
Another  important feature is that wherever two bosons 
in the same spin state (except $m=0$), say located at $\Vr$ and
$\Vr^\prime$, approach each other, $\Psi^b_N$ 
vanishes as  $\sim (\Vr-\Vr^\prime)^2$, as implied in Table~\ref{tab:energy}.
Readers who are not interested in the technical details
may skip now to the end of the paper 
for other interesting physical properties.

\paragraph{Projective construction: a theoretical technique.}

Unlike the usually studied, weakly interacting dilute limit, a Bose
gas of strong repulsion cannot be treated by conventional perturbation
theory. How to derive the groundstate wavefuntion is obviously
challenging. Furthermore, even if one has such a poposed wavefunction 
as the $\Psi^b_N$ above, it is still difficult to extract the low
energy excitation properties, since it is a highly `entangled' state.  
The analogue of Gross-Pitaevskii equation does not exist yet. To make
some progress towards such a difficult task, we borrow the `projective
construction' method from quantum Hall studies, 
which is a standard, successful approach in
parallel with the Laughlin wavefunction approach 
(see Ref. \cite{Wen:99} and references therein).  
In this spirit,  we introduce two spin-${1\over 2}$ fermions,
$\psi_{a\al}$, each with two (physical) spins $\al=\up, \down$ and two
fictitious ``colors'' labelled by $a, b$. The color is necessary to
furnish a minimal spin-1 bosonic representation  at every space-time point.
The boson can then be  represented by
\be
\phi_{m}(\Vr)=  \psi_{a\al}(\Vr)\psi_{b\bt}(\Vr)
\eps_{ab}\CC^m_{\al\bt} \,.
\label{eq:phi=}
\ee
Note that the color degree of freedom is non-physical. All physical
states or operators are required 
invariant under a {\it local} \sucolor\ transformation:
$
\psi_{a\al}(\Vr) \rightarrow W(\Vr)_{ab}\psi_{b\al}(\Vr) \,,
$ 
where $W(\Vr)$ is an \sucolor\ matrix. One can quickly check that
the boson operator $\phi_m$ defined in (\ref{eq:phi=}) is indeed
a color singlet, perfectly invariant under above transformation. 
Eq.~(\ref{eq:phi=}) allows us to construct physical boson many-body
wave function $|\Psi^b\>$
from the unphysical fermion many-body wave function$|\Psi^f\>$ by
projecting it into the
``color'' singlet sector. In a mathematical equation, that means
\be \textstyle
\Psi^b(\Vr_1, m_1; \Vr_2, m_2; ...) = \avg{0|\prod_i \phi_{m_i}(\Vr_i)
|\Psi^f} \,,
\label{eq:Psib=f}
\ee
whose explicit form is precisely $\Psi^b_N$ of
Eq.~(\ref{eq:Psib=cluster}). 

The relationship (\ref{eq:Psib=f})  suggests that the low energy
effective theory
of our spin-1 boson system can
either be described in terms of the boson operator $\phi_m$, or equivalently,
in terms of the fermion operator $\psi_{a\al}$.
Unfortunately, for a strongly correlated 2D system, there is no known rigorous
way to derive the effective theory from a microscopic Hamiltonian
like (\ref{eq:Hboson}).  One usually first writes down a
most natural form of it on symmetry grounds, checks its stability
against interactions in low energy limit, and finally
compares it with experiments.  In this spirit,  the 
effective theory for the
state $\Psi^b(\Vr_1, m_1;\Vr_2, m_2; ...)$  is described,
in the fermion description, by a theory of independent fermions coupled to
color SU(2) gauge fields. The gauge field is denoted as
$A_\mu\equiv {1\over 2} \tau^l a_\mu^l$, $l=1,2,3$, 
where ${\tau^l}$ are the Pauli
matrices generating the \sucolor\ algebra.
The gauge fields are introduced to project out the
unphysical colored excitations.\cite{Wen:99}  
The effective theory is then
\bea
\CL  &=& \textstyle
i \psi^{\dag}_{a\al} (D_0)_{ab} \psi_{b\al}
 + {1 \over 2M}\psi^{\dag}_{a\al}(\VD\cdot\VD)_{ab} \psi_{b\al} \nn \\
 && +\mbox{all symmetry allowed interactions} \,,
\label{eq:su2-eff}
\eea
where $(D_\mu)_{ab}\equiv \del_{ab}\pt_\mu - i (A_\mu)_{ab}$ (for notation,
see \cite{note:mu})
are the covariant derivatives. From the
fermion effective theory, we can study various fermion states, which, after
the projection (\ref{eq:Psib=f}), lead to various physical boson states.

To see which fermion states are likely to appear as the ground state, we need
to consider the interactions between the fermions.  Interactions can be either
originated from the boson-boson interactions or dynamically generated by gauge
interactions.  As an example of non-Abelian gauge theory, the study of QCD
\cite{Schafer:98} shows that the Yang-Mills gauge fluctuations can generate a
strong attractive interaction between quarks due to the instanton effect,
which leads to quark confinement. In our case, the gauge fields are used to
mediate a strong attractive interaction between color-opposite fermions, since
by definition (\ref{eq:phi=}) the gauge interaction is supposed to  bind two
fermions locally into a colorless boson.  Therefore, the strong SU(2) gauge
interaction naturally leads to color-singlet Cooper pairing.    

Let us consider two simplest
color-singlet parings: 
\bea 
 && \textstyle
\avg{\psi_{a\al}(\Vr)\psi_{b\bt}(0)} \nonumber \\
&=& \textstyle
\left\{
\begin{array}{ll}
\eps_{ab} \, \CC^{\bar{m}}_{\al\bt}\, R_s(|\Vr|) \,, 
\quad & \mbox{($s$, spin triplet)} \\
\eps_{ab} \, \eps_{\al\bt}\, R_p(|\Vr|)\, (x +iy)  \,,
& \mbox{($p_{x+iy}$, spin singlet)}  	
\end{array}\right. \label{eq:pair}
\eea
where $\bar{m}$ can be $0$ or $1$.
Both $s$- and $p_{x+iy}$-wave states produce a full gap on the Fermi surface.
$R_{s,p}(|\Vr|)$ are complex functions of $|\Vr|$, 
generically expected to monotonically fall off exponentially 
at large distance.  

\paragraph{$s$-wave pairing: conventional BEC phases.} Conventional BEC phases
are easily recovered through the $s$-wave paring channel.  In the
limit of strong confinement, 
$R_s(|\Vr|)$ becomes a delta function $\sim \del(\Vr)$.

\subparagraph{Polar condensate.} This is a special kind of spin nematic state.
Fermion confinement occurs in spin-1, $m=0$ channel. Here, the pairing
in Eq.(\ref{eq:pair}) reduces to 
$
\avg{\psi_{a\al}(\Vr)\psi_{b\bt}(\Vr)} = 
\sqrt{\rho} e^{i\vphi} \eps_{ab} \CC^{m=0}_{\al\bt}
$
where one may think of $\rho$ related to the condensed boson density.  The
resulting state is nothing but the so-called polar condensate with
$\avg{\phi_{m=0}}\neq 0$.  This state breaks both \uch and
\suspin\ invariance, and was considered to be favored  if
the spin interaction is antiferromagnetic ($V_2 \gtrsim 
V_0$) \cite{Ho-Yip:98+00}. 

\subparagraph{Ferromagnetic condensate.} 
This case is the same as the polar
condensate except that the fermions are confined into the $m=1$ (or
equivalently $-1$) channel.  The order parameter becomes
$
\avg{\psi_{a\al}(\Vr)\psi_{b\bt}(\Vr)} = 
\sqrt{\rho} e^{i\vphi} \eps_{ab} \CC^{m=1}_{\al\bt}
$
which corresponds to the ferromagnetic condensate. Like the polar
condensate, it breaks both \uch and \suspin\ invariance.
This state is presumably favored if the spin interaction is ferromagnetic,
$V_2 \lesssim V_0$.

\paragraph{$p_{x+iy}$-state: topological superfluid.}
The spin-disordered
$p_{x+iy}$-wave pairing energetically competes with the polar
condensate for a large positive $V_2$ (see Table.~\ref{tab:energy}).
Assuming the state to exist, its low energy effective theory 
can be routinely constructed,
simply based on the broken (physical) symmetries without relying on
microscopic details. We find the effective Lagrangian of the state,
\bea 
&& \textstyle
\int_\Vr \left\{
\psi^{\dag}_{a\al} i(D_0)_{ab} \psi_{b\al} +
{1\over 2M} \psi^{\dag}_{a\al}{\VD^2_{ab}} \psi_{b\al} \right\} +
\label{eq:su2-pwave}
\\
&& \textstyle
\int_{\Vr\Vr^\prime} \left[
\psi^\dag_{a\al}(\Vr)\Del(\Vr,\Vr^\prime) \eps_{\al\bt}
[e^{i\int_{\Vr^\prime}^\Vr \boldsymbol{A}\cdot d\boldsymbol{l}} i\tau^2]_{ab} 
\psi^\dag_{b\bt}(\Vr) + \mbox{h.c.}\right] \,, \nn 
\eea
where $\Del(\Vr,\Vr^\prime)$ is the $p$-wave pairing wavefunction.
In the momentum space it becomes $\Del(\Vk) \equiv {\Del_0\over
2}(k_x+ ik_y)$.   The theory is  evidently invariant under the global \suspin\
and local \sucolor\ transformations. However, since the $p_{x+iy}$-wave
condensate (see Eq.~(\ref{eq:pair})) spontaneously breaks the charge U(1), 
the phase fluctuations of $\Del$ gives rise to the usual
(gapless) superfluid mode.
The charge sector therefore is rather conventional.  In contrast, we shall
show that the spin sector has excitations with fractional spin and statistics.

Let us focus on the spin sector of the effective theory
(\ref{eq:su2-pwave}), by regarding $\Del$ as a fixed non-dynamical field. 
To see the symmetry more clearly, we introduce four $2$-component  
spinors, each uniting a pair of particle and hole operators:
\be 
\eta_{a\al}\equiv {\psi_{a\al} \choose \eps_{ab}\eps_{\al\bt} 
\psi^\dag_{b\bt}} \,,
\qquad (a,b=1,2;\, \al,\bt=\up,\down)\,.  \label{eq:eta=}
\ee
In terms of $\eta$'s, the effective action in
momentum space becomes 
\bea
I_{\rm eff} &=& \textstyle 
- {1\over 2} \int d^3k \eta^\dag_{a\al}(k) 
[\CD(k^\mu-A^\mu)]_{a\al;b\bt}
\eta_{b\bt}(k) \,, \label{eq:Ieff} \\
\CD(k) &=& \hat{\tau}^0\otimes
\tau^0\otimes[-\omega + \vec{h}(\Vk)\cdot\vec{\sig}]
\,,\  k^\mu\equiv(\omega, \Vk)\,, \ \label{eq:Dmatrix}\\
\vec{h}(\Vk) &\equiv & 
\left(\Del_0 k_y\,, \ \Del_0 k_x \,, \  {\Vk^2/ 2M}-\mu \right)
\,. 
\label{eq:hk}
\eea
The $\CD(k)$ is better viewed as an $8\times 8$ matrix 
in an orthogonal basis 
$\{\hat{\tau}^0=\Bid,\hat{\tau}^l\}_{\rm spin} \otimes 
\{\tau^0=\Bid,{\tau}^l\}_{\rm color} 
\otimes \{\sig^0=\Bid,{\sig}^l\}$ ($l=1,2,3$)  
with the
$\sig$'s acting on the ``Nambu'' spinor space (\ref{eq:eta=}).
A careful reader may recall 
$A_\mu= {1\over 2} {\tau}^l{a}^l_\mu$.
  
The spectrum of fermionic excitations is determined by the pole of the
inverse of $\CD(k)$ at $A_\mu =0$, 
$
E(\Vk) = \sqrt{\left(\Vk^2/2M -\mu\right)^2 +\Del_0^2} \,.
$
The fermions are fully gapped, so we integrate out them to find the effective
action of gauge fields. 
To do so, we have chosen a background field
gauge in (\ref{eq:Ieff}) 
such that $A_\mu=\mathit{Const}$ but non-commuting, $[A_\mu,
A_\nu]\neq 0$. The approach we adopted here 
is  standard in non-Abelian gauge field theory. Since the action
(\ref{eq:Ieff}) is quadratic in fermion fields, the gauge effective
Lagrangian is exactly given by
\be \textstyle
\CL_{\rm eff}[A] =  -{i\over 2(2\pi)^3}\int d^3k
\Tr\ln\CD(k-A) \label{eq:LeffA}
\ee
where the `$\Tr$' is over the internal space of $\vec{\hat{\tau}}
\otimes \vec{\tau} \otimes
\vec{\sig}$. 
With the $\CD$ matrix given in (\ref{eq:Dmatrix}), a straightforward
calculation gives 
\be \textstyle
\CL_{\rm eff}[A] = {Q\over 4\pi}\eps^{\mu\nu\rho} \tr\{A_\mu \pt_\nu A_\rho
-{2i \over 3} A_\mu A_\nu A_\rho\}  + \cdots \,, \label{eq:LCS}
\ee
where $Q$ is a (topological) winding number ($Q=1$ \cite{note:Q=1})
and the
`$\cdots$' stands for higher derivatives, including a Maxwell
term. Now the `\tr' in (\ref{eq:LCS}) is over only the color gauge
indices.
The effective theory of gauge bosons is thus 
a level $1$ (non-Abelian) SU(2) Chern-Simons theory. All gauge excitations
gain a dynamically generated topological mass (i.e., gapped). The gauge
interaction becomes short ranged. So the
effective theory we found in (\ref{eq:su2-pwave}) is stable.

\paragraph{Physical properties of the
$p_{x+iy}$-state. } 
After obtaining the low energy effective theory, we are ready to study the
measurable properties of the state. First the $p_{x+iy}$-state is a superfluid
which does not break spin rotation symmetry. The only gapless excitation is
the superfluid mode described by the phase of $\Delta$.  All spin excitations
have a finite energy gap. Due to the Chern-Simons term, the $SU(2)$ gauge field
is not confining. Thus the excitations described by $\psi$ (or $\eta$) have a
finite energy gap (instead of infinite energy gap). Those excitations carry
spin-1/2 and one half of boson charge! They also have a semion statistics
({\it i.e.}, the statistical angle is $\theta=\pi/2$, right between boson and
fermion), as implied by the level-1 $SU(2)$ Chern-Simons terms.  The spin
rotation symmetry implies that $\<\phi_m\>=0$ for all $m$. 
However, two-boson and
three-boson operators both can have finite expectation values. In
short distance, 
$\avg{\phi_m(z_1)\phi_{m^\prime}(z_2)} \sim -4\sqrt{3}
(z_1-z_2)^2  
C^{00}_{mm^\prime}$ 
and 
$\avg{\phi_{m_1}(z_1)\phi_{m_2}(z_2) \phi_{m_3}(z_3)} \sim 8\sqrt{2}
(z_1-z_2) (z_2-z_3)(z_3-z_1) \eps_{m_1m_2m_3}$ where $z\equiv x+iy$. 
The spin-disordered superfluid has an unusual off-diagonal long range
order.  We see that the minimum vortex has one unit of quantized
vorticity.  Our $p_{x+iy}$ spin-disordered state breaks the parity and time
reversal symmetry. The total angular momentum of the ground state
is $\hbar$ per
boson.  (Note such a  total angular momentum  is equal to the total angular
momentum of usual boson superfluid with one vortex at its center.) Thus,
spinning the bosons may help to create the $p_{x+iy}$ spin-disordered superfluid.

\paragraph{Edge excitations}
In the $\eta$ bases, the mean-field fermion Hamiltonian 
(described by (\ref{eq:Ieff}) with $A_\mu$ being set to zero)
contains four identical $2\times
2$ blocks: $H=\vec h(\Vk)\cdot \vec \sigma$. 
In each block, the function $\vec h(\Vk)$ defines a mapping from the
$\Vk$-space to $S^2$ with a winding number $1$. This non-trivial winding
number leads to a unit Hall conductance.\cite{TKN8205,ASS8351} Since each
block contributes a unit Hall conductance, it leads to an edge state similar
to the one from $\nu=1$ quantum Hall state,\cite{H8285} as required by gauge
invariance.\cite{Wedge,H9397,V9792} Such an edge state can be described by one
chiral fermion field $\lambda_{a\al}$ (one for each block).  
Therefore, if we ignore the $SU(2)$ gauge fluctuations, the mean-field gapless
edge excitations of our $p_{x+iy}$-state
are described by the following effective theory:
$\lambda^\dagger_{a\al} i(\partial_t-v \partial_x) \lambda_{a\al}$.
Only two of the four  $\lambda$'s are  independent because $\eta_{a\al}$ (see
(\ref{eq:eta=})) are Majorana fermions and each $\lambda_{a\al}$ is obtained
as a linear combination of the two components of spinor $\eta_{a\al}$.
Therefore, the mean-field edge state contains only {\it two} independent
branches of chiral fermions.  Obviously the mean-field edge effective theory
has \sucolor\ and \suspin\ symmetries generated by $\tau^l \otimes \hat{\tau}^0$
and $\tau^0 \otimes \hat{\tau}^l$, $l=1, 2, 3$, respectively.  Having the
\sucolor\ $\times$ \suspin\ symmetry and a central charge $c=2$ ({\it i.e.}, two
branches of chiral fermions), we find that the mean-field edge state is
described by  \sucolor\ $\times$\suspin\ Kac-Moody current algebra of level 1.
After including the \sucolor\ gauge fluctuations to go beyond the mean-field
theory, the edge effective theory becomes
\begin{equation}
\label{edge}
 \lambda^\dagger_{a\al} i \left[
(\partial_t-iA_0)-v (\partial_x-iA_x) \right]_{ab} \lambda_{b\al} \,.
\end{equation}
The effect of \sucolor\ gauge fields is to remove the \sucolor\ sector of the
Kac-Moody algebra from the low energy spectrum.\cite{Wedge}
Thus the physical edge state
of the $p_{x+iy}$-state is described by a level-1 \suspin\ Kac-Moody algebra.
The physical edge state is identical to the right moving sector of 
spin-1/2  chain.
Despite the finite gap in the bulk, the spin excitation is gapless at the
edge. The operator that creates the gapless spin-1/2 quasiparticle on the edge
is given by the spin-1/2 primary field $V_\al(x,t)$ in the \suspin\ Kac-Moody
algebra which has a scaling dimension 1/4.  The quasiparticle propagator has a
form $\< V_\al(x,t) V_\al(0)\>\sim(x-vt)^{-1/2}$. The boson operator $\phi_m$
on the edge becomes the spin-1  primary field which is the spin current
operator on the edge.  The boson propagator on the edge is given by $\<
\phi_m(x,t) \phi^\dagger_m(0)\>\sim(x-vt)^{-2}$. (The boson propagator is
short ranged in the bulk due to the finite spin gap.) This will lead to a
non-linear I-V curve $I\propto |V|^2 V$ for boson tunneling between two edges.
The spin-1/2 quasiparticles can tunnel between two edges separated by a
bulk $p_{x+iy}$-state. The tunneling I-V curve has a form $I\propto |V|^{-1}V$
in the weak tunneling limits. Finally, we briefly mention that
the polar condensate has gapless spin excitations
in the bulk whereas the $p_{x+iy}$ spin-disordered
superfluid has a gapless spin excitation only 
at the edge. A dramatic difference
can be seen in the spin susceptibility by NMR experiments.

\paragraph{Conclusions}
A two-dimensional boson gas in ultra-cold alkali atomic 
systems can be strongly correlated.  A
spin-1 boson system can have a very interesting spin-disordered 
superfluid state,
which carries a non-trivial topological order.\cite{Wtoprev}  Such possibility
is interesting, since they might exhibit some of the spin liquid phases that
have been long theoretically speculated in the context of high $T_c$
superconductors but never been clearly identified by experiments.  We
believe that the alkali atomic gases may provide the first controlled
laboratory to check those speculated theories and enrich our understanding of
the strongly correlated systems.  In fact the $p_{x+iy}$-state is closely
related to the chiral spin state.\cite{wwz}  The two states have the same
bulk effective theory described by level-1 \sucolor\ Chern-Simons
theory,\cite{W90} and the same edge effective theory described by level-1
\suspin\ Kac-Moody algebra.\cite{Wedge} The spin sector of the two
states are identical.


This work is supported by NSF grants DMR-0123156 and DMR-9808941.
W.V.L. is also  supported  by funds
provided by the U.S. Department of Energy (DOE) under cooperative
research agreement \#DF-FC02-94ER40818.

\bibliographystyle{apsrev} 
\bibliography{becsl}

\begin{thebibliography}{10}
\expandafter\ifx\csname bibnamefont\endcsname\relax
  \def\bibnamefont#1{#1}\fi
\expandafter\ifx\csname bibfnamefont\endcsname\relax
  \def\bibfnamefont#1{#1}\fi
\expandafter\ifx\csname url\endcsname\relax
  \def\url#1{\texttt{#1}}\fi
\expandafter\ifx\csname urlprefix\endcsname\relax\def\urlprefix{URL }\fi
\providecommand{\bibinfo}[2]{#2}
\providecommand{\eprint}[2][]{\url{#2}}

\bibitem{Dalfovo:99}
\bibinfo{author}{\bibfnamefont{F.}~\bibnamefont{Dalfovo}},
  \bibinfo{author}{\bibfnamefont{S.}~\bibnamefont{Giorgini}},
  \bibinfo{author}{\bibfnamefont{L.~P.} \bibnamefont{Pitaevskii}},
  \bibnamefont{and}
  \bibinfo{author}{\bibfnamefont{S.}~\bibnamefont{Stringari}},
  \bibinfo{journal}{Rev. Mod. Phys.}
  \textbf{\bibinfo{volume}{71}}(\bibinfo{number}{3}), \bibinfo{pages}{463}
  (\bibinfo{year}{1999}).

\bibitem{Fisher:88}
\bibinfo{author}{\bibfnamefont{D.~S.} \bibnamefont{Fisher}} \bibnamefont{and}
  \bibinfo{author}{\bibfnamefont{P.~C.} \bibnamefont{Hohenberg}},
  \bibinfo{journal}{Phys. Rev. B} \textbf{\bibinfo{volume}{37}},
  \bibinfo{pages}{4936} (\bibinfo{year}{1988}).

\bibitem{Ketterle:01pre}
\bibinfo{note}{A. G\"orlitz, et al., cond-mat/0104549.}

\bibitem{Hansel:01}
\bibinfo{author}{\bibfnamefont{W.}~\bibnamefont{H\"ansel}},
  \bibinfo{author}{\bibfnamefont{P.}~\bibnamefont{Hommelhoff}},
  \bibinfo{author}{\bibfnamefont{T.~W.} \bibnamefont{H\"ansch}},
  \bibnamefont{and} \bibinfo{author}{\bibfnamefont{J.}~\bibnamefont{Reichel}},
  \bibinfo{journal}{Nature} \textbf{\bibinfo{volume}{413}},
  \bibinfo{pages}{498} (\bibinfo{year}{2001}).

\bibitem{Ohmi:98}
\bibinfo{author}{\bibfnamefont{T.}~\bibnamefont{Ohmi}} \bibnamefont{and}
  \bibinfo{author}{\bibfnamefont{K.}~\bibnamefont{Machida}},
  \bibinfo{journal}{J. Phys. Soc. Jpn.} \textbf{\bibinfo{volume}{67}},
  \bibinfo{pages}{1822} (\bibinfo{year}{1998}).

\bibitem{Ho-Yip:98+00}
\bibinfo{note}{T. L. Ho, Phys. Rev. Lett. {\bf 81}, 742 (1998); T. L. Ho and S.
  K. Yip, {\it ibid} {\bf 84}, 4031 (2000).}

\bibitem{Law:98}
\bibinfo{author}{\bibfnamefont{C.~K.} \bibnamefont{Law}},
  \bibinfo{author}{\bibfnamefont{H.}~\bibnamefont{Pu}}, \bibnamefont{and}
  \bibinfo{author}{\bibfnamefont{N.~P.} \bibnamefont{Bigelow}},
  \bibinfo{journal}{Phys. Rev. Lett.} \textbf{\bibinfo{volume}{81}},
  \bibinfo{pages}{5257} (\bibinfo{year}{1998}).

\bibitem{Demler-Zhou:01pre}
\bibinfo{note}{E. Demler and F. Zhou, cond-mat/0104409; F. Zhou,
  cond-mat/0108473}.

\bibitem{Wen:99}
\bibinfo{author}{\bibfnamefont{X.-G.} \bibnamefont{Wen}},
  \bibinfo{journal}{Phys. Rev. B} \textbf{\bibinfo{volume}{60}},
  \bibinfo{pages}{8827} (\bibinfo{year}{1999}), \bibinfo{note}{and references
  therein}.

\bibitem{note:mu}
\bibinfo{note}{We adopt the Greek index notation:
  $(t,\Vr)=x^\mu=\eta^{\mu\nu}x_\nu$ with $\mu,\nu=0,1,2$ and a metric
  $\eta^{\mu\nu} =(-,+,+)$.}

\bibitem{Schafer:98}
\bibinfo{note}{For a recent review, see, T. Sch\"afer and E. V. Shuryak, Rev.
  Mod. Phys. {\bf 70}, 323 (1998).}

\bibitem{note:Q=1}
\bibinfo{note}{To calculate the winding number, one first variationally
  differentiates both sides of (\ref{eq:LeffA}) three times with respect to the
  gauge field ${\del^3\cdots / \del A_\mu^l \del A_\nu^m \del A_\rho^n}$, sends
  all $A_\mu^l\rightarrow 0$, and then contracts it with $\eps_{\mu\nu\rho}
  \eps^{lmn}$. One finds Q is given by a topological invariant, $ Q= {(1/8\pi)}
  \eps_{ij} \int d^2\Vk \, \hat{h} \cdot ({\pt\hat{h}/ \pt k_i} \times
  {\pt\hat{h}/\pt k_j})$, $\hat{h} \equiv {\vec{h}/ |\vec{h}|}\,. $ Inserting
  the $\vec{h}(\Vk)$ functions defined in (\ref{eq:hk}) gives $Q=1$.}

\bibitem{TKN8205}
\bibinfo{author}{\bibfnamefont{D.~J.} \bibnamefont{Thouless}},
  \bibinfo{author}{\bibfnamefont{M.}~\bibnamefont{Kohmoto}},
  \bibinfo{author}{\bibfnamefont{M.~P.} \bibnamefont{Nightingale}},
  \bibnamefont{and} \bibinfo{author}{\bibfnamefont{M.}~\bibnamefont{den Nijs}},
  \bibinfo{journal}{Phys. Rev. Lett.} \textbf{\bibinfo{volume}{49}},
  \bibinfo{pages}{405} (\bibinfo{year}{1982}).

\bibitem{ASS8351}
\bibinfo{author}{\bibfnamefont{J.}~\bibnamefont{Avron}},
  \bibinfo{author}{\bibfnamefont{R.}~\bibnamefont{Seiler}}, \bibnamefont{and}
  \bibinfo{author}{\bibfnamefont{B.}~\bibnamefont{Simon}},
  \bibinfo{journal}{Phys. Rev. Lett.} \textbf{\bibinfo{volume}{51}},
  \bibinfo{pages}{51} (\bibinfo{year}{1983}).

\bibitem{H8285}
\bibinfo{author}{\bibfnamefont{B.~I.} \bibnamefont{Halperin}},
  \bibinfo{journal}{Phys. Rev. B} \textbf{\bibinfo{volume}{25}},
  \bibinfo{pages}{2185} (\bibinfo{year}{1982}).

\bibitem{Wedge}
\bibinfo{note}{X.-G. Wen, Int J. Mod. Phys. {\bf B6}, 1711 (1992); Phys. Rev.
  {\bf B43}, 11025 (1991).}

\bibitem{H9397}
\bibinfo{author}{\bibfnamefont{Y.}~\bibnamefont{Hatsugai}},
  \bibinfo{journal}{Phys. Rev. Lett.} \textbf{\bibinfo{volume}{71}},
  \bibinfo{pages}{3697} (\bibinfo{year}{1993}).

\bibitem{V9792}
\bibinfo{author}{\bibfnamefont{G.}~\bibnamefont{Volovik}},
  \bibinfo{journal}{Pis\'ma ZhETF} \textbf{\bibinfo{volume}{66}},
  \bibinfo{pages}{492} (\bibinfo{year}{1997}).

\bibitem{Wtoprev}
\bibinfo{author}{\bibfnamefont{X.-G.} \bibnamefont{Wen}},
  \bibinfo{journal}{Advances in Physics} \textbf{\bibinfo{volume}{44}},
  \bibinfo{pages}{405} (\bibinfo{year}{1995}).

\bibitem{wwz}
\bibinfo{author}{\bibfnamefont{X.-G.} \bibnamefont{Wen}},
  \bibinfo{author}{\bibfnamefont{F.}~\bibnamefont{Wilczek}}, \bibnamefont{and}
  \bibinfo{author}{\bibfnamefont{A.}~\bibnamefont{Zee}},
  \bibinfo{journal}{Phys. Rev. B} \textbf{\bibinfo{volume}{39}},
  \bibinfo{pages}{11413} (\bibinfo{year}{1989}).

\bibitem{W90}
\bibinfo{author}{\bibfnamefont{X.-G.} \bibnamefont{Wen}},
  \bibinfo{journal}{Mean Field theory of Spin Liquid States and Topological
  Orders}  (\bibinfo{year}{1990}), \bibinfo{note}{http://dao.mit.edu/\~\/wen}.

\end{thebibliography}

\end{document}